\documentstyle[aps,epsfig,twocolumn,prb]{revtex}

\title{Unified Model, and Novel Reverse Recovery Nonlinearities, of the Driven Diode Resonator} 
\author{Renato Mariz de Moraes \thanks{\ Present address: Electrical Engineering Department, University of California at Santa Cruz, Santa Cruz, CA}
 and Steven M. Anlage  \\
Center for Supercondutivity Research, Department of Physics, University of Maryland,\\ 
College Park, MD, 20742-4111, USA (e-mail: anlage@squid.umd.edu).}

\begin{document} 
\maketitle 
\begin{abstract} 
We study the origins of period doubling and chaos in the driven series resistor-inductor-varactor 
diode (RLD) nonlinear resonant circuit.  We find that resonators driven at frequencies much 
higher than the diode reverse recovery rate do not show period doubling.  Models of chaos 
based on the nonlinear capacitance of the varactor diode display a reverse-recovery-like 
effect, and this effect strongly resembles reverse recovery of real diodes.  We find for the 
first time that in addition to the known dependence of the reverse recovery time on past 
current maxima, there are also important nonlinear dependencies on pulse frequency, 
duty-cycle, and DC voltage bias.  Similar nonlinearities are present in the nonlinear 
capacitance models of these diodes.  We conclude that a history-dependent and nonlinear 
reverse recovery time is an essential ingredient for chaotic behavior of this circuit, and 
demonstrate for the first time that all major competing models have this effect, either 
explicitly or implicitly.   Besides unifying the two major models of RLD chaos, our work 
reveals that the nonlinearities of the reverse recovery time must be included for a complete 
understanding of period doubling and chaos in this circuit.
\end{abstract}

\newpage  
 
\section{Introduction} 

\begin{figure}[h] 
\centerline{\epsfxsize=6.5cm \epsfbox{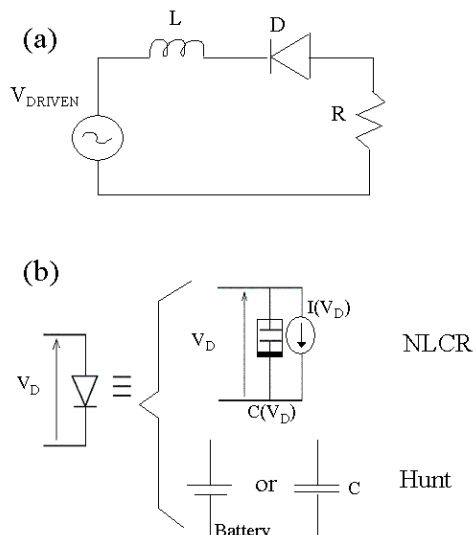}}     
\caption{Schematic diagrams for the (a) driven RLD circuit, and (b) diode models.  The 
nonlinear capacitance and nonlinear resistance (NLCR) implementation is shown in the upper 
part of (b), while a representation of the Hunt, $et al.$ model is shown in the lower part.  
In the latter case, the diode is replaced with a battery while it is conducting, and a fixed 
linear capacitor at all other times.}
\label{Fig1}
\end{figure}

The nonlinear dynamics of the p-n junction has been a subject of intense interest since the 
dawn of chaos.\cite{Linsay1981,Testa1982,Hunt-Testa1982,Rollins+Hunt1982,Brorson1983,Matsumoto1984,vanBuskirk1985,Su1989}  Experimental work has focused on the damped driven 
nonlinear oscillator formed by a resistor, inductor, and varactor diode (RLD circuit) 
connected in series (see Fig.\ \ref{Fig1}(a)).  This is the simplest passive circuit that 
displays period doubling and chaos as a function of sinusoidal driving amplitude and frequency.  The dynamics of this circuit can, in principle, be understood through the details of the hole and electron charge distributions in the vicinity of the p-n junction.\cite{Millman}  However, many simplified models of p-n junction charge dynamics have been put forward in terms of nonlinear lumped-element approximations for the varactor (Fig.\ \ref{Fig1}(b)).  One reason for this approach was the interest in constructing low-dimensional maps to describe the "universal" nonlinear dynamics experimentally observed in the circuit.\cite{Linsay1981,Testa1982,Rollins+Hunt1982,Matsumoto1984} 

There are three general types of nonlinearity present in the driven RLD circuit.  The first 
is the nonlinear current-voltage characteristic of the diode, thought by most researchers to 
be unimportant for chaos, because its absence from models results in only small changes in 
the calculated bifurcation diagrams.\cite{Brorson1983,Carroll2002}  A second contribution 
comes from the large and exponentially nonlinear forward-bias capacitance associated with 
the junction diffusion capacitance.\cite{Brorson1983}  In this case, the diode is modeled 
as a parallel combination of a nonlinear resistor and nonlinear capacitor (Fig.\ \ref{Fig1} (b)).  
It has been proposed that period doubling will occur when the capacitance reaches 
approximately four times its zero bias value because the resonant frequency of the RLD circuit 
will drop to half its low-amplitude value.\cite{vanBuskirk1985}  This is thought to be the 
first step in a period doubling cascade to chaos.  A third contribution to nonlinearity comes 
from the finite time-scale diffusive dynamics of charge in the p-n junction and the 
associated "memory" of previous forward-current maxima.\cite{Rollins+Hunt1982}  After the 
diode has been forward-biased and switched off, it takes some time for the minority carriers 
to diffuse back across the junction, allowing the diode to conduct for a period known as the 
reverse-recovery time, ${\tau}_{RR}$.  The "memory" of previous forward bias currents built 
into this non-equilibrium charge distribution has been proposed as the main source of chaos 
in the driven RLD circuit.\cite{Rollins+Hunt1982,Su1989}

Lumped element models of the driven RLD circuit fall into two distinct classes: those 
relying on the nonlinear capacitance and ignoring explicit history effects (Fig.\ \ref{Fig1}(b) NLCR), 
and those explicitly based on history effects in the junction and ignoring the nonlinear 
capacitance (Fig.\ \ref{Fig1}(b) Hunt).  Both kinds of model attempt to identify the essential 
physics that is responsible for period doubling and chaos.  It is remarkable that both classes 
of models reproduce intricate detail of the return maps and bifurcation diagrams of the driven 
RLD circuit.  Historically, the two classes of models were assumed to be mutually exclusive 
and contradictory in their approach to describing chaotic dynamics in the driven RLD 
circuit.\cite{ADH1983}  One purpose of this paper is to partially reconcile these two 
classes of models, and to show that they both include some essential physics of this rich 
nonlinear problem.  We also demonstrate that both models lack important effects that have 
been ignored up to this point.  These effects are measured here for the first time in the 
context of nonlinear dynamics, and are essential for a complete understanding of the behavior 
of this circuit.

In this paper we present new measurements and simulations of reverse-recovery times in a 
variety of diodes and resonant circuits.  Section \ref{sec_exp} discusses the many time 
scales of the driven RLD problem and illustrates the importance of the reverse recovery 
time in the nonlinear dynamics of the circuit.  Section \ref{sec_phase} demonstrates that 
the nonlinear capacitance models are endowed with a history-dependent reverse-recovery-like 
effect.  Section \ref{sec_dc} is a summary of our experimental results of the nonlinear 
dependence of the reverse recovery time on circuit parameters, and ends with a dramatic 
illustration of how existing models fail to explain key data.  Sections \ref{sec_obs} 
and \ref{sec_conc} review our contributions to a deeper understanding of the RLD circuit 
and demonstrate that a simple nonlinear capacitor model shows all of the essential 
features of reverse-recovery-like behavior.

Experiments were carried out in a manner similar to that described in Ref. \cite{Moraes2002}.  
The circuits were shielded inside a metallic box and excited through 50 $\Omega$ transmission 
lines by either an HP 33120A synthesized source (up to 15 MHz) or an HP 83620B microwave 
synthesizer (above 10 MHz).  The response of the circuits was measured with either a 
Tektronix TDS3052 oscilloscope, or a Tektronix 494P spectrum analyzer.  Low capacitance 
and high impedance probes were used to measure voltages in the circuits.  SPICE simulations 
showed that the probes had a minimal impact on the circuit waveforms, whereas it was found 
that low impedance capacitive probes strongly perturbed the circuits.  Bifurcation diagrams 
were constructed by collecting single-shot traces from the oscilloscope and recording the 
data with a LabView program on a PC.

\section{Time Scales} 
\label{sec_exp} 

The history of research on the driven RLD circuit is long and varied.  Most papers on this 
subject study only one diode and choose a circuit with a resonant frequency in a convenient 
range (typically at audio frequencies).  The experiments have been carried out in widely 
disparate regions of the parameter space available for the circuit (the parameters include 
driving frequency, amplitude, and duty-cycle, R and L values, minority carrier lifetime, dc 
voltage bias, etc.).\cite{Klinker1984,Perez1985,Ast1988,Baxter1990,Tanaka1996}  Only recently 
has effort been made to assemble a unifying picture of the origins and generality of period 
doubling and chaos in this important circuit.\cite{Carroll2002}  Our purpose in this section 
is to measure and model different diode circuits in a wide range of parameter space to 
identify the important and essential physics responsible for the nonlinear dynamics of 
the driven RLD circuit as a step in creating such a unified picture.

It is important to first understand the four major time/frequency scales involved in 
this problem, because the dynamics of the circuit depends critically on the relative 
values of these time scales.  The first is the low-amplitude resonant frequency of the 
passive RLD circuit: f$_0$ = 1/[2$\pi$(LC$_j$)$^{1/2}$], where C$_j$ is the zero-bias 
junction capacitance of the varactor diode.  It is well established that complex dynamics 
are most easily observed for driving frequencies f in the vicinity of this resonant frequency.  Another important time scale in the diode is the reverse recovery time: ${\tau}_{RR}$.  This is a measure of the time required for minority carriers injected across the junction in the diode to move back to the other side of the junction (or recombine) once the driving force has been removed.  One effect of this finite time scale is "reverse recovery" in which a forward-biased diode will not shut off immediately when the driving signal is reversed, but will continue to conduct for a time on the order of ${\tau}_{RR}$.  Su, Rollins and Hunt proposed that the diode actually acts like a battery during this time.  In addition, they proposed that ${\tau}_{RR}$ retained a memory of the two or three previous maximum forward current values through the diode.\cite{Su1989}  

There are several RC time constants in the problem, associated with the junction 
capacitance and the internal resistance of the diode, and between the junction 
capacitance and the external resistor, R (note that the parasitic resistance of 
the inductor is much less than R for the circuits considered here).  These time 
scales are usually quite short ($\sim$ ns), but can be important when the capacitance 
is large.  Finally, the L/R time scale is generally long and does not directly influence 
the diode dynamics.  The interplay of these time scales, along with the period of the 
sinusoidal driving signal, yield a very rich parameteric landscape for nonlinear dynamics.

It should be mentioned that three of these time scales are strongly nonlinear.  The 
resonant frequency f$_0$ and the RC time constants are a nonlinear function of driving 
amplitude and DC voltage bias on the varactor junction, and we show for the first time 
below, the reverse recovery time is a nonlinear function of drive frequency, amplitude, 
duty-cycle, DC bias, and load resistance.\cite{Millman}  Most of these nonlinearities 
have not been investigated in the past with respect to the resulting nonlinear dynamics.  
These nonlinearities greatly enrich the dynamics and challenge our understanding of the 
essential physics responsible for nonlinear dynamics in the simple driven RLD circuit.

	To illustrate the interplay of time scales in this problem, consider the sinusoidally 
driven resonant RLD circuit shown in Fig.\ \ref{Fig1}(a).  The low-amplitude resonant 
frequency of the circuit is determined by the diode junction capacitance C$_j$ as well as 
the lumped element inductor, L.  The reverse recovery time ${\tau}_{RR}$ can be measured by 
methods discussed below.  We studied four different diodes with very different values 
of C$_j$ and ${\tau}_{RR}$.  In all cases we find the following.  First, when the 
low-amplitude resonant frequency of the circuit f$_0$ is on the order of 1/${\tau}_{RR}$, 
period doubling and chaos are found for a substantial driving frequency range f between 
about 0.2 f$_0$ and 2 f$_0$ (see Table I).  Similar results are obtained when the inductance 
is changed to make the resonant frequency equal to about one tenth of 1/${\tau}_{RR}$ (in 
agreement with Ref. \cite{Carroll2002}).  However, when the inductance is changed to 
make f$_0$ about ten times larger than 1/${\tau}_{RR}$, there is a substantial change in 
the circuit behavior (Table I).  One of the diode circuits shows only period doubling and 
no chaos, while another displays nothing but period-one behavior.  The other diode circuits 
show period doubling and chaos, but over a more restricted range of amplitude and driving 
frequency parameter space.  Finally, when the circuits are modified to 
have f$_0$ $\sim$ 100/${\tau}_{RR}$, we do not observe period doubling or chaos at all 
over the equivalent range of parameters.

	When the diode is driven at frequencies much higher than the reverse recovery 
time, the charge carriers near the p-n junction do not have a chance to influence the 
circuit behavior or contribute additional nonlinear behavior.  Clearly the reverse 
recovery time plays an important role in the nonlinear dynamics of this circuit.  However, 
very little prior experimental work on RLD dynamics has explored the interplay of all the 
time scales in this problem.

\section{The Storage (${\tau}_{s}$) and Reverse Recovery (${\tau}_{RR}$) Time Scales} 
\label{sec_phase} 

Hunt and collaborators claimed that the reverse recovery time is the key feature that 
generates chaos in the driven RLD circuit because it is a function of the peak current 
flow in previous cycles, thus introducing a memory effect into the dynamics.  In the 
paper by Rollins and Hunt \cite{Rollins+Hunt1982} a simple model of reverse recovery 
was introduced for RLD dynamics, and turned into a 1D noninvertible map.  Although this 
gave qualitative agreement with experiment, it was criticized for being inconsistent with 
the measured attractor.\cite{Brorson1983}  Later, Su, Rollins and Hunt \cite{Su1989} 
included three or more previous cycles of the forward bias current to determine their 
model ${\tau}_{RR}$.  They also noted that the recovery time depends on the junction 
characteristics and the external series resistance.  We show here that this model is a 
good starting point, but fails to include essential nonlinearities.

\begin{figure}[h] 
\centerline{\epsfxsize=10cm \epsfbox{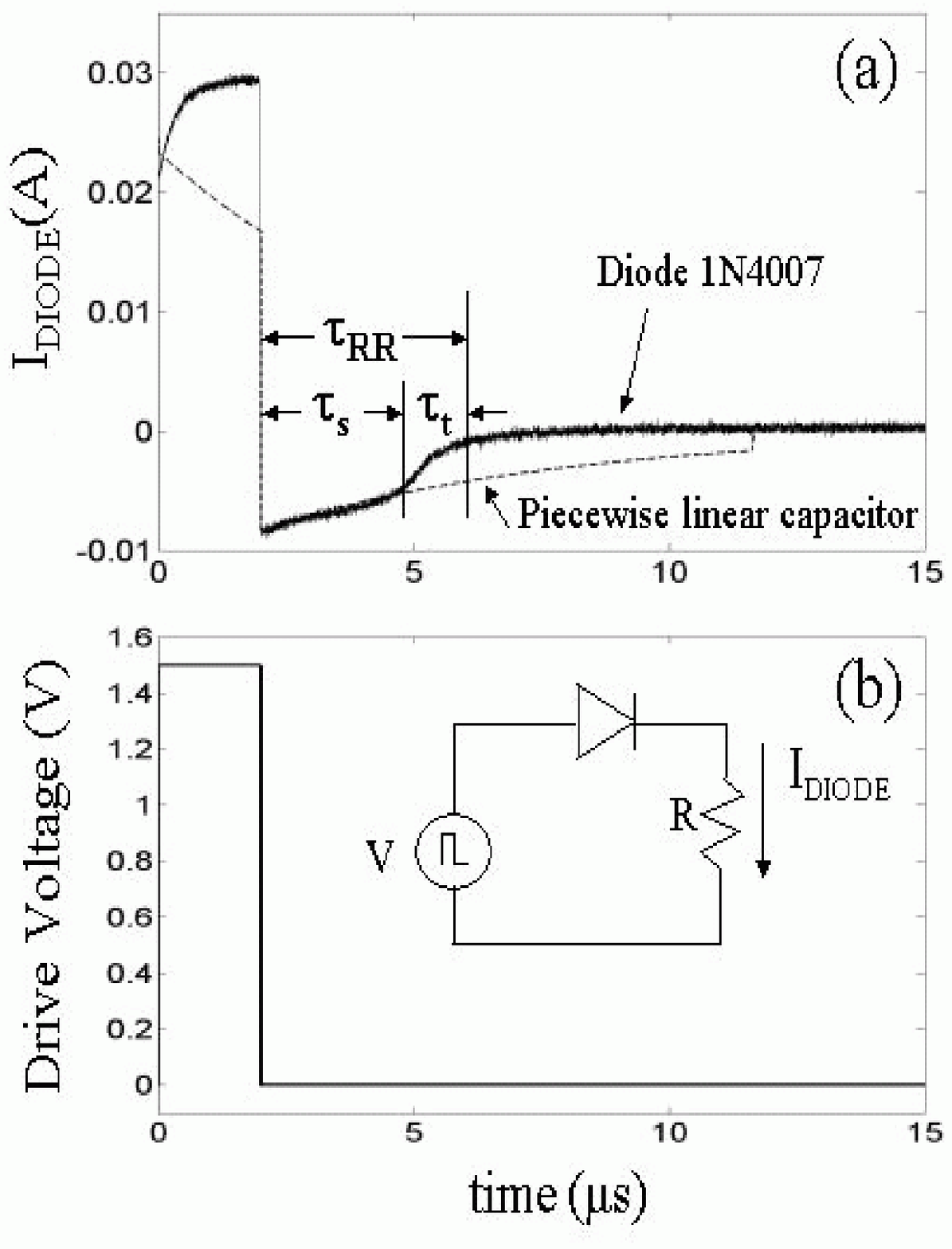}}     
\caption{(a) Current flowing through the diode during and after the pulse excitation 
shown in (b), demonstrating the reverse recovery effect for a 1N4007 diode (solid line), 
and the discharge effect on a piecewise linear capacitor (PLC) model in place of the 
diode (dashed line). For the diode circuit R=25 $\Omega$, while for the PLC \cite{Matsumoto1984} R = 60 $\Omega$, C$_1$ = 0.1 $\mu$F, C$_2$ = 400 pF, and E$_0$ = 0.1 V  (see Fig. 3(b)).  These values for the PLC model are used throughout the paper.  (b) Shows the drive voltage applied to both circuits and the inset shows the circuit schematic. }
\label{Fig2}
\end{figure}

\begin{figure}[h] 
\centerline{\epsfxsize=10cm \epsfbox{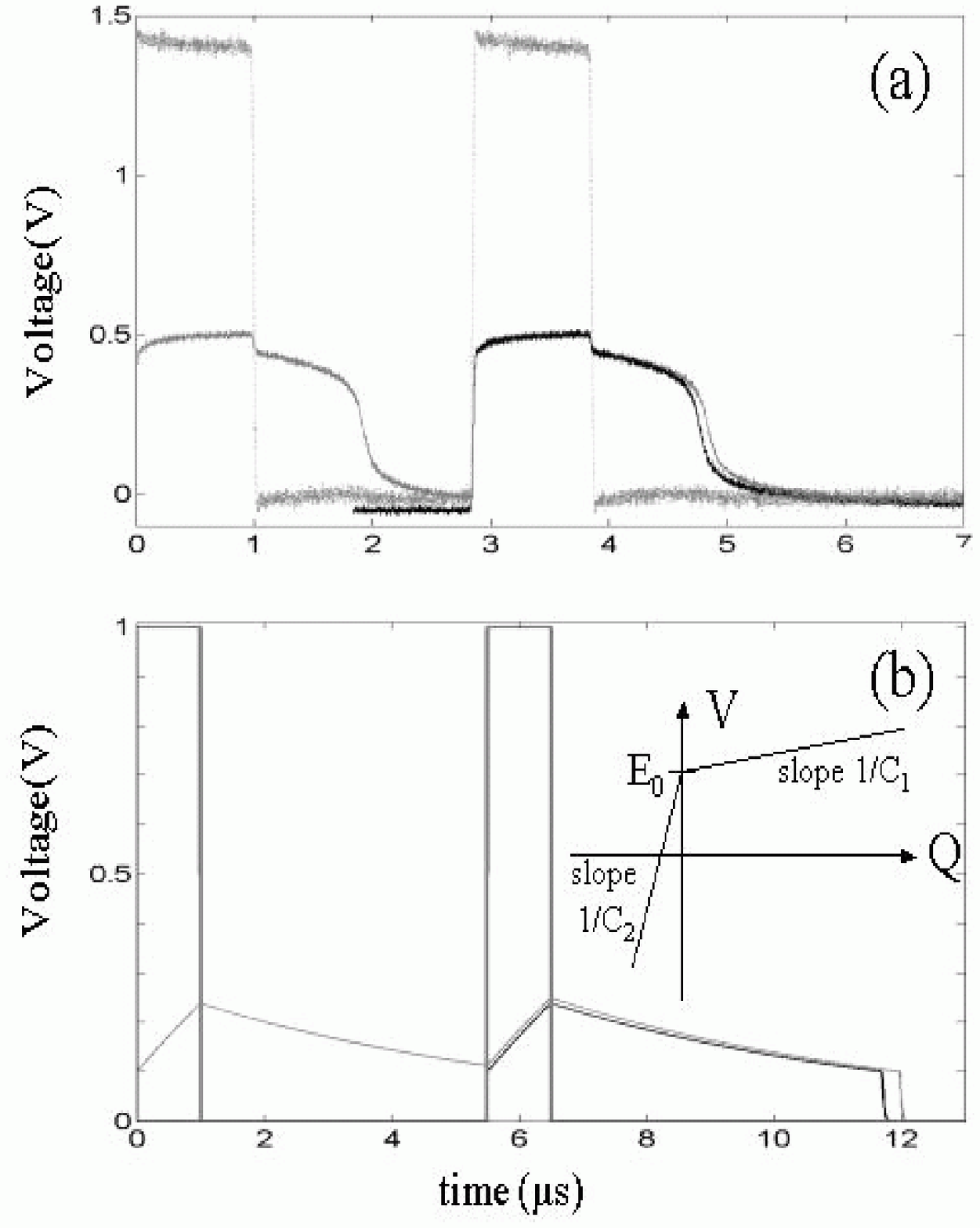}}     
\caption{Illustration of reverse-recovery history dependence through measurements 
of the voltage drop on the device (diode or PLC) after either one or two pulse 
excitation (taller curves) using the circuit in the inset of Fig. 2(b).  The lighter 
curves show double pulse exposure and the darker curves show a single pulse 
exposure. (a) Diode 1N4007 and R=1 k$\Omega$, (b) PLC model (implemented in SPICE) 
using the same circuit parameters as in Fig. 2.  The PLC voltage-charge curve is 
shown in the inset.  Note that in both cases the device requires a longer time to 
recover after the second pulse compared to just one pulse.}
\label{Fig3}
\end{figure}

	Figure \ref{Fig2} illustrates the reverse recovery effect in a diode.  The circuit 
shown in the inset of Fig.\ \ref{Fig2}(b) was employed to drive the diode with a single 
square pulse input.  After the driving signal was removed, the diode continued to conduct 
for a time scale defined as the reverse recovery time, ${\tau}_{RR}$.  This time consists 
of two contributions: the storage time ${\tau}_{s}$, and the transition time ${\tau}_{t}$, 
with ${\tau}_{RR}$ = ${\tau}_{s}$ + ${\tau}_{t}$.\cite{Millman}  We shall present results 
below (all data in Figs. \ \ref{Fig3} - \ref{Fig7} are taken with the circuit shown in 
Fig.\ \ref{Fig2}(b)) mainly on the storage time because, in general, this time scale 
dominates the reverse recovery time and its nonlinear behavior.

	Figure \ref{Fig3} illustrates the history dependence of the storage and reverse 
recovery times.  In this case we compare single pulse recovery data like that shown 
in Fig.\ \ref{Fig2} to the same diode subjected to two consecutive pulses (Fig. \ref{Fig3}(a)).  The storage time is longer for the diode that has been subjected to two pulses, mainly because it was not finished recovering from the first pulse before the second arrives.  This simple history dependence of the recovery time is the basis for the Hunt and Rollins treatment of complex RLD dynamics.\cite{Rollins+Hunt1982,Hunt+Rollins1984}

	All other models of RLD dynamics do not consider an explicit history dependence 
of the junction dynamics of the diode.  They are based on the nonlinear behavior of 
the capacitor used to approximate the reactive electrical behavior of the 
diode (Fig.\ \ref{Fig1}(b), NLCR).  As we shall show here, in fact all of these 
other models do indeed have a history-dependence of a reverse-recovery-like time 
that naturally comes from the charge storage on the capacitor part of the 
varactor diode model.

\subsection{Piece-wise Linear Capacitor (PLC) Model} 
\label{sec_enhanc} 

To illustrate our point, we shall choose the simplest model of the nonlinear capacitance 
in the varactor.  In this case, one can use a piece-wise linear approximation for the 
capacitance of the varactor, with just two linear pieces.  This model for the varactor 
capacitance was considered by several groups\cite{Matsumoto1984,ADH1984,Bez1994} and 
shown to produce period doubling and chaos when part of an R-L-PLC series circuit driven 
by a sinusoidal signal.  The inset to Fig.\ \ref{Fig3}(b) shows that the PLC switches at 
potential E$_0$ from a large value C$_1$ for higher bias, to a small value C$_2$ for 
lower bias.   We use this model to illustrate that it has implicitly included a 
history-dependent storage time.  Fig.\ \ref{Fig2}(a) shows a SPICE simulation of the 
diode current (dashed line) using the PLC model for the varactor in the circuit shown 
in the inset of Fig.\ \ref{Fig2}(b), where the diode is replaced with the PLC.  It clearly 
displays a "storage time" and a "transition time", just like a real 
diode.\cite{Carrollfootnote}  The "storage time" (${\tau}_s$) in this case corresponds 
to the RC time constant of the capacitor discharging into the external resistance of 
the circuit while it has a large capacitance C$_1$.  The "transition" part of the 
signal occurs when the device switches to its smaller capacitance value C$_2$, producing 
a much shorter RC time constant that discharges the capacitor very quickly.  Thus the 
discharge behavior of the capacitor mimics the effects of minority charge carrier 
diffusion across the junction.  As we shall see, the details of this discharge behavior 
are remarkably similar to the diffusive dynamics of the minority charges in real diodes.

	We have found that a variety of models based on nonlinear capacitance of the diode 
show a history dependence of the RC storage and "reverse recovery" 
times.  Fig.\ \ref{Fig3}(b) illustrates the history-dependence of the RC discharge 
time in the PLC model.  All models of RLD dynamics that we have investigated have a 
history-dependent reverse recovery effect coming from the nonlinear capacitance.  Our 
conclusion is that a finite reverse recovery time is indeed an important concept for 
understanding chaos in this circuit.  Besides making this point in this paper, we also 
wish to explore the nonlinear and history-dependent behavior of this recovery time and 
how it influences the generic behavior of the driven RLD circuit.

\section{Nonlinear Behavior of the Storage Time}
\label{sec_dc}

The storage time and its history dependence are clearly important ingredients in 
any model of nonlinear dynamics of the driven RLD circuit.  However, it is not 
widely appreciated that the storage time is itself a strongly nonlinear function 
of many parameters in the problem.  These parameters include the amplitude of the 
forward bias current, the circuit resistance R, the driving frequency, the duty 
cycle of the drive, and the DC voltage bias on the junction.  The purpose of this 
section is to present experimental evidence and models of these nonlinearities and 
to discuss their significance for the occurrence of period doubling and chaos in 
the parameter space of these circuits.

First is the forward current amplitude dependence of the storage time ${\tau}_s$.  Hunt 
and Rollins pointed out that the reverse recovery time of a diode increases as the 
amplitude of the forward bias current through the diode increases.  The approximate 
expression used to describe this effect was;\cite{Rollins+Hunt1982,Hunt+Rollins1984}
 
\begin{equation}
{\tau}_{RR} = {\tau}_{m} ( 1 - exp(-|I_{max}|/I_c)) ,
\label{Eq1}
\end{equation}
where I$_{max}$ is the most recent maximum forward current, ${\tau}_{m}$ and I$_c$ are 
lifetime and current scales particular to each diode.  A more exact expression for the 
storage time ${\tau}_{s}$ can be derived from analysis of charge dynamics in the p-n 
junction;\cite{Neudeck}

\begin{equation}
erf(\sqrt{\frac {{\tau}_{s}}{{\tau}_{m}}}) = \frac {1}{1+I_R / I_{max}},
\label{Eq2}
\end{equation}
that can be approximated as;
	 								
\begin{equation}
{\tau}_{s} \approx {\tau}_{m} ln( 1 + \frac {I_{max}}{I_R} ) ,
\label{Eq3}
\end{equation}
where I$_R$ is the reverse current through the diode during the storage phase.  All of 
these forms basically say that the storage and reverse recovery times will increase as 
the amplitude of the forward current increases.  This is because more minority charge 
will be pushed across the junction by the larger current and, because of the diffusive 
dynamics of the carriers, there will be a longer delay in getting back to the equilibrium 
charge configuration.

\begin{figure}[h] 
\centerline{\epsfxsize=6.5cm \epsfbox{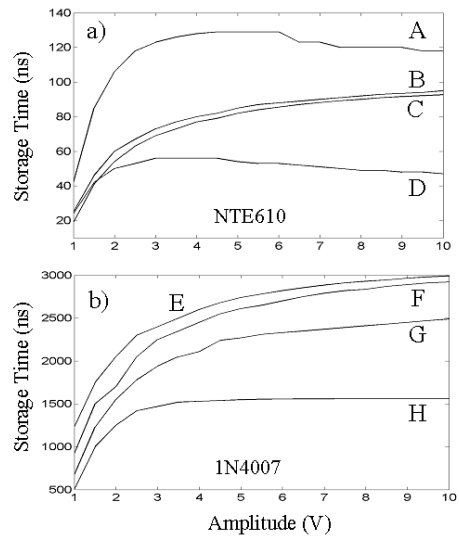}}     
\caption{Squared wave amplitude dependences: a) Amplitude dependence of the storage time for NTE610 diode with (A) R=1 k$\Omega$ 
experiment, (B) R=1 k$\Omega$ PSpice model, (C) R=25 $\Omega$ PSpice, and (D) R=25 $\Omega$ 
experiment.  b) Amplitude dependence of the storage time for 1N4007 diode with (E) 
R=1 k$\Omega$ PSpice, (F) R=25 $\Omega$ PSpice, (G) R=1 k$\Omega$ experiment, and (H) 
R=25 $\Omega$ experiment.  Measurements and simulations were done using the inset circuit 
shown in Fig. 2(b) where the driving voltage was a square wave shape (with no DC bias) of frequency 
20 kHz (as the recovery time is constant for low frequencies). The peak-to-peak voltage
was twice the amplitude.}
\label{Fig4}
\end{figure}

\begin{figure}[h] 
\centerline{\epsfxsize=6.5cm \epsfbox{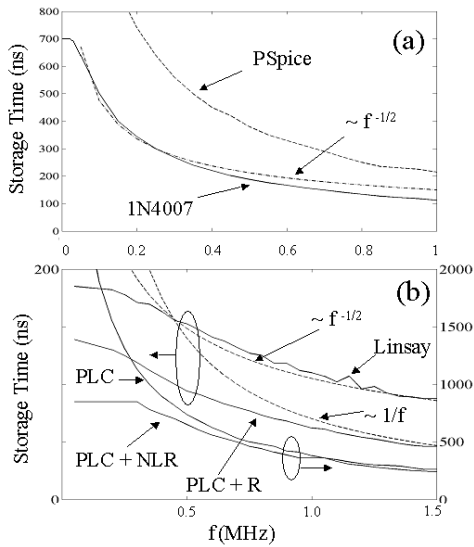}}     
\caption{Frequency dependence of storage time (for diodes) or discharging effect 
(for PLC cases). Here we use a square wave of amplitude 1.5 V (3 V peak-to-peak), no DC bias, 50 \% duty 
cycle. (a) Diode 1N4007 and R=25 $\Omega$. Solid line is the experimental result, dashed line is the 1N4002 PSpice model, and 
dashed-dotted line is a 1/f$^{1/2}$ dependence. (b) PLC SPICE simulations of discharging 
storage time using parameters from Ref. 4.  The nonlinear resistor (NLR) uses the I-V curve 
of Ref. 3, while the PLC+R uses a R = 10 $\Omega$ parallel resistor.  The Linsay model is a 
PSpice simulation using the diode model of 
Ref. 1 with R = 25 $\Omega$.  Also shown as dashed lines are 1/f and 
1/f$^{1/2}$ dependencies.}
\label{Fig5}
\end{figure}

	Data and PSpice simulations for the forward current dependence of the storage time 
in two different diodes is shown in Fig.\ \ref{Fig4}.  (Note that PSpice simulations use 
standardized semi-empirical models of the diode behavior, while our SPICE simulations of 
the PLC use the model shown in the inset of Fig.\ \ref{Fig3}(b) with parameters given in 
the caption of Fig.\ \ref{Fig2}.)  The circuit in the inset of Fig.\ \ref{Fig2}(b) was 
used to perform these measurements and simulations.  For the data in Figs.\ \ref{Fig4} and  \ref{Fig5} the driving signal is a periodic square wave with an equal time spent at the positive and negative amplitude voltage values (50\% duty cycle).  The storage time was measured as a 
function of the driving signal amplitude, proportional to the forward current through the 
diode.  The data and simulations all show an increase of the storage time with increased 
forward bias current.  In some cases the storage time saturates, consistent with the 
approximate expression Eq. (1), and in other cases it continues to rise slowly, consistent 
with the more sophisticated treatment in Eqs. (2) and (3).

	It has also been noted that the value of the resistance R in the circuit of 
Fig.\ \ref{Fig2}(b) (and therefore Fig.\ \ref{Fig1}(a)) will also influence the storage 
and reverse recovery times.\cite{Su1989,Hunt+Rollins1984}  This is illustrated 
in Fig.\ \ref{Fig4}.  When the load resistance  R is changed from 25 $\Omega$ to 1 k$\Omega$, 
the storage time increases for both diodes (NTE610 and 1N4007) in both experiment and 
simulation.  This is expected because the reverse recovery process involves charge transport 
across the junction and through the rest of the circuit.  A larger external resistance in the 
circuit will imply a smaller reverse recovery current, thus extending the time for the 
minority charge to re-establish its equilibrium configuration.  Note that a similar effect 
will occur in models of nonlinear capacitance.  In that case the discharge time of the 
capacitor is an RC time constant, which clearly grows with external 
resistance R.\cite{Roulston}

\begin{figure}[h] 
\centerline{\epsfxsize=10cm \epsfbox{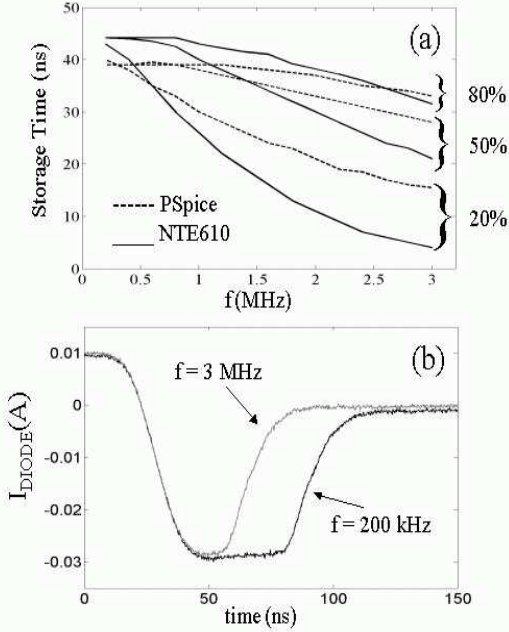}}     
\caption{(a) Frequency and duty cycle dependence (20\%, 50\%, and 80\%) of the storage time 
for a NTE610 diode with R=25 $\Omega$, a square wave amplitude of 1.5 V (3 V peak-to-peak),
and no DC bias. Solid lines are experimental 
results and dashed lines are for a calculation using the FMMV2101 PSpice Zetex model. (b) 
Experimental time series of diode current for the 50\% duty cycle square wave case (from item (a)) for frequencies of 
200 kHz and 3 MHz, illustrating the decrease in reverse recovery time as the frequency 
increases.}
\label{Fig6}
\end{figure}

	The nonlinearities of the storage time dependence on forward bias current and 
resistance have been noticed in the past.  However, we are not aware of any group 
noting the following nonlinearity in the reverse recovery time.  When the circuit 
in the inset of Fig.\ \ref{Fig2}(b) is used with a repetitive square wave pulse 
sequence, one finds a strong dependence of the reverse recovery time on the pulse 
frequency.  We consider the case where the pulses remain sufficiently far apart that the 
diode appears to fully recover from the previous pulse before the next arrives.  As the 
rate of pulsing increases, the recovery time dramatically decreases.  The effect is 
illustrated in Figs. \ \ref{Fig5} and \ \ref{Fig6}.  Fig. \ \ref{Fig6}(b) shows recovery 
traces from an NTE610 diode at two pulse frequencies.  With a pulse repetition rate of 
200 kHz, the storage time is about 40 ns, while at a repetition rate of 3 MHz it has 
decreased to less than 20 ns.  Figs. \ \ref{Fig5} and \ \ref{Fig6} show that in general 
the storage time is constant at low frequencies, and then begins to fall roughly 
like f$^{-1/2}$ with increasing frequency.  This behavior is seen in both experiment 
and PSpice simulation of the 1N4002, NTE610, and Linsay\cite{Linsay1981} model diodes.

	From examination of charge dynamics in the p-n junction it is known that the 
diffusion capacitance C$_D$ is a function of frequency $\omega$ for sinusoidal 
driving signals;\cite{Millman}
	 		
\begin{equation}
C_D = \cases{\tau_{m} g/2, &  $\omega \tau_{m} \ll $ 1 \cr
                 g \sqrt{\tau_{m} \over {2 \omega}}, & $\omega \tau_{m} \gg $ 1 \cr} 
\label{Eq4}
\end{equation}       

where g is the differential conductance of the diode g = dI/dV.  The f$^{-1/2}$ frequency 
dependence of the capacitance comes from the diffusive nature of charge transport in the 
junction region.\cite{Millman}  The diffusion capacitance diminishes with increasing 
frequency beyond f $\sim$ 1/2$\pi {\tau}_m$, resulting in less charge stored in the p-n 
junction for a given forward bias current.  This in turn implies that less minority carrier 
charge must be moved across the junction at high frequencies, resulting in a shorter storage 
and reverse recovery time.  This is consistent with the results presented in Figures \ \ref{Fig5} and \ref{Fig6} for two rather different diodes (1N4007 and NTE610).  

	One possible consequence of the frequency-dependent storage time is the persistence 
of period doubling behavior to unexpectedly high frequencies in the driven RLD circuit.  
As the driving frequency increases, the diffusion capacitance and storage time decrease, 
effectively increasing both the resonant frequency f$_0$ and 1/${\tau}_{RR}$.  This 
nonlinearity essentially helps to maintain the condition 
f $\sim$ f$_0$ $\sim$ 1/${\tau}_{RR}$ (where period doubling and chaos is 
most easily observed) over an extended range of driving frequencies.  Thus period 
doubling and chaotic behavior should be more robust in the driven RLD circuit than one 
might naively expect.  
  
	Fig.\ \ref{Fig6}(a) illustrates the duty-cycle dependence of the storage time 
and its frequency dependence.  The duty cycle is defined as the ratio of the time spent at the positive amplitude voltage value to the total period of the periodic square pulse signal.  As the duty cycle increases, there is greater overlap 
between adjacent pulses, causing the enhancement of the reverse recovery time, as 
originally noted by Hunt and collaborators and summarized in Eqs. (1)-(3).  Again we see 
very consistent behavior in both experiment and PSpice simulations.

\begin{figure}[h] 
\centerline{\epsfxsize=10cm \epsfbox{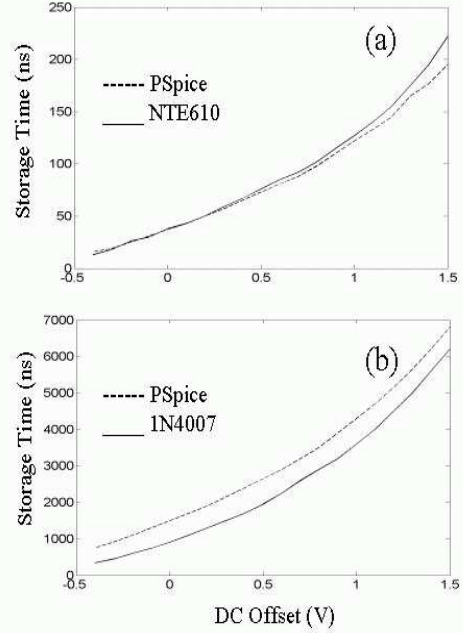}}     
\caption{DC voltage offset dependence of the storage time for two diodes. 
 R=25 $\Omega$, amplitude of 1.5 V (3 V peak-to-peak), f=20 kHz, 50 \% duty cycle.  (a) Solid line is the experimental result on NTE610 
diode and the dashed line is for a PSpice simulation using the FMMV2101 Zetex model. 
(b) Solid line is for experimental result on 1N4007 
diode and dashed line is for PSpice simulation using FMMV2101 Zetex model.}
\label{Fig7}
\end{figure}

	Finally, we have noted a strong DC voltage bias dependence of the storage time.  The 
data and simulations on two different diodes is shown in Fig.\ \ref{Fig7}.  For forward bias 
the storage time increases dramatically, while for negative bias it decreases.  The range of 
storage time variation is more than a factor of 10 for a 2 volt swing in DC bias voltage.  A 
forward-biased diode will have a broader region of minority charge distribution, forcing the 
carriers to travel greater distances.  This in turn increases the recovery time of the diode.  
Similarly, a reverse biased diode is able to sweep the minority carriers across the junction 
more quickly with the additional driving force provided by the reverse DC bias.  This strong 
nonlinearity of the recovery time has not been discussed in the nonlinear dynamics 
literature, to our knowledge.

\begin{figure}[h] 
\centerline{\epsfxsize=10cm \epsfbox{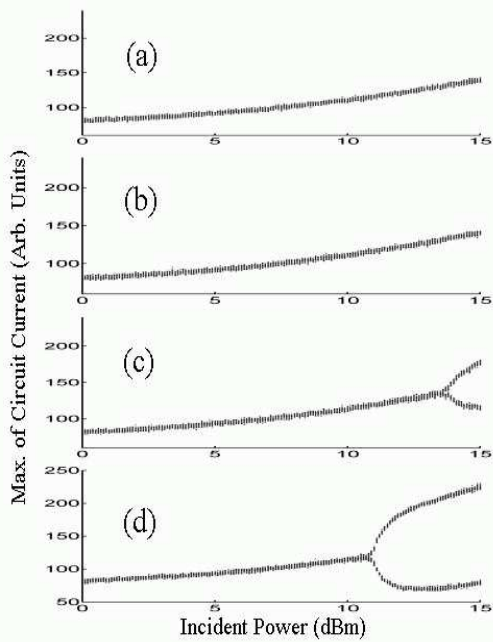}}     
\caption{Bifurcation diagrams for RLD circuit (Fig. 1(a)) using an NTE610 diode with 
R=25 $\Omega$, f=29 MHz, and L=10 $\mu$H (note that f is greater than f$_0$ = 12.3 MHz). 
(a) DC Offset = +0.2 V, (b) DC Offset = 0 V, (c) DC Offset = -0.4 V, 
(d) DC Offset = -0.6 V.}
\label{Fig8}
\end{figure}

	The DC bias nonlinearity can have a significant effect on the bifurcation diagram of 
the driven RLD circuit.\cite{Maksimov,Kim1997}  As a DC bias is applied, the storage time 
and differential capacitance change, and the relationship between f$_0$, 1/${\tau}_{RR}$, 
and the driving frequency f will change.  As discussed above, this can influence whether or not chaos is observed 
in the circuit.  Here we present an example of the complex dynamics that can result.  
Figure \ref{Fig8} shows a case where the driving frequency f is larger than 1/${\tau}_{RR}$ and
 about two to three times the RLD circuit resonant frequency f$_0$ at zero DC bias.  In this 
bifurcation diagram we see only period-1 behavior for driving signal power up to +15 dBm for 
both zero and positive DC bias.  However, for negative bias, we see period doubling set in at 
about +14 dBm for a -0.4 V bias and about +11 dBm for a -0.6 V bias.  The prevailing NLC 
models cannot explain this observation because a negative bias moves the diode away from 
the nonlinear part of the C(V) curve.  In this case we believe that the increase 
in 1/${\tau}_{RR}$ and increase in f$_0$ due to V$_{DC}$ create a situation where 
f $\sim$ f$_0$ $\sim$ 1/${\tau}_{RR}$, and the circuit displays period doubling at 
frequencies where ordinarily it only shows linear response.  

	It should be noted that period doubling and chaos can be observed in models of the 
driven RLD circuit even in the apparent absence of reverse recovery.\cite{Hunt-Testa1982,Carroll2002,Kim1988}  
However, it is not clear to what extent these models may have discharge-time effects that 
mimic the behavior of reverse recovery in diodes. 

\section{Discussion}
\label{sec_obs}

To make further comparisons between the nonlinear capacitor (NLC) and reverse-recovery time 
models of RLD dynamics, we wish to further consider the simplified PLC model of the NLC.  The 
PLC model contains the essential physics of all NLC models, and is very easy to implement and 
understand.  Figure \ref{Fig2}(a) (dashed line) shows that the PLC model of the diode 
(Fig. \ \ref{Fig3}(b), inset) clearly shows an effective delay in "turning off" after a large 
forward current has been applied to the device.  This delay is a combination of two RC time 
constants.  The first, or "storage" part of the response is associated with the RC discharge 
of the PLC during its high-capacitance phase.  As the voltage drop on the PLC decreases, it 
reaches the point (V = E$_0$) where the capacitance switches to its smaller value, resulting 
in a much more rapid discharge, resembling the transition phase of reverse recovery.   The 
simulation in Fig.\ \ref{Fig2}(a) (dashed line) for the PLC qualitatively resembles the data 
on the 1N4007 diode.  (Note that this simulation is not intended to fit the precise behavior 
of the 1N4007 diode, but to simply reproduce the essential behavior.)

	The history dependence of the discharge time for the PLC model is illustrated in 
Fig.\ \ref{Fig3}(b)  (all simulations of the PLC model are done in SPICE).  The discharge 
time is clearly enhanced after the second pulse of a two pulse sequence, compared to just 
a single pulse.  This result is similar to that in Ref. \cite{Hunt+Rollins1984} and 
Fig.\ \ref{Fig3}(a) on real diodes.  There is also strong frequency dependence to the 
discharge time of the PLC model, as illustrated in Fig.\ \ref{Fig5}(b).  The PLC capacitor 
model alone shows a discharge time that varies as 1/f.  When a linear resistor is added in 
parallel with the PLC, the discharge time drops more slowly with frequency (PLC +R in 
Fig.\ \ref{Fig5}(b)).  Finally, when a nonlinear resistor (NLR) is added in parallel to 
the PLC, using the ideal diode IV curve of Linsay \cite{Linsay1981}, the discharge time 
shows behavior remarkably similar to the reverse recovery time of a real diode.    

	In summary, the discharge behavior of the simple PLC model (when implemented in the 
circuits of interest) is remarkably similar to the reverse recovery behavior of real diodes.  
We find that SPICE simulations of other more sophisticated NLC models give very similar 
results.  We conclude that all NLC models will show a reverse-recovery-like effect.  It 
appears that the discharge of the NLC/PLC model is essentially the same as the diffusion 
of charge carriers out of the p-n junction.  Hence this important ingredient of nonlinear 
physics is common to the two major models of nonlinear dynamics of the driven nonlinear 
resonator.  However, our results demonstrate that all of these models lack key additional 
nonlinear properties of the reverse-recovery time.
 
\section{Conclusions} 
\label{sec_conc} 

We maintain that reverse recovery is a key effect producing chaos in the sinusoidally driven 
RLD circuit.  All nonlinear capacitor models of this circuit have a behavior that mimics 
reverse recovery.  Both the reverse-recovery and nonlinear capacitor models have memory 
built into them through a charge storage mechanism.  The details of these models are 
different, but the results for circuit dynamics are basically the same.  We find that 
it is not useful to make the distinction that either the nonlinear capacitance or the 
reverse recovery on minority carriers is the cause of chaos, because they can have similar 
effects on the circuit dynamics.  Finally, the reverse recovery time in real diodes is a 
strongly nonlinear function of forward bias current, driving frequency and DC bias.  A 
detailed understanding of chaos in the RLD circuit must include the reverse recovery effect 
and all of its nonlinearities.

\section{Acknowledgements} 
\label{sec_ack} 
This work was supported by STIC through the STEP Program, and by the Department of Defense 
MURI Program under AFSOR Grant F496 200 110 374. R. M. de Moraes was also supported by 
CAPES-Brazil. We acknowledge helpful discussions with Tom Carroll, and early work on the 
circuit was done by Sang-Ho Bok.

\onecolumn

\begin{table}[tbp] 
\caption{\label{var}
Survey of results for period doubling and chaos on driven RLD circuits with four different 
diodes.  The reverse recovery time ${\tau}_{RR}$ was measured at 20 kHz (where it is 
frequency independent) with V$_{DC}$ = 0, a 1.5 V amplitude pulse (3 V peak-to-peak), 50 \% duty cycle, and R = 25 $\Omega$.  
The zero-bias junction capacitance C$_j$ was measured with an Agilent 4285A LCR meter.  
The resonant frequencies f$_0$ were changed by adjusting the L value in the RLD circuit, 
while R was fixed at 25 $\Omega$.  The search for period doubling and chaos took place 
over the range f $\sim$ 0.1 f$_0$ to f $\sim$ 10 f$_0$.  We were unable to measure a 
finite reverse recovery time (i.e. ${\tau}_{RR}$ $<$ 10 ns) in the 1N4148 fast-recovery 
diode due to the bandwidth limitations of our oscilloscope.  Thus we were unable to 
establish the condition f $\sim$ f$_0$ $\sim$ 1/${\tau}_{RR}$ for that diode.} 
\vspace{0.2in}
\begin{tabular}{cccccc}  
\hline
\hline
Diode   & ${\tau}_{RR}$ (ns)  & C$_j$ (pF)& Results with  & Results with  & Results with  \\ 
&&& f$_0$ $\sim$ 1/${\tau}_{RR}$ & f$_0$ $\sim$ 10/${\tau}_{RR}$ & f$_0$ $\sim$ 100/${\tau}_{RR}$ \\
\hline
1N5400 & 7000 & 81 & Period doubling and  & Period doubling and & No period doubling \\
&&& chaos for  & chaos for  & or chaos \\
&&& f/f$_0$ $\sim$ 0.11 - 1.64 & f/f$_0$ $\sim$ 0.16 - 1.76 &\\

1N4007 & 700 & 19 & Period doubling and  & Period doubling and & No period doubling \\
&&& chaos for  & chaos for  & or chaos \\
&&& f/f$_0$ $\sim$ 0.13 - 2.0 & f/f$_0$ $\sim$ 0.23 - 1.3 &\\

1N5475B & 160 & 82 & Period doubling and  & No period doubling & No period doubling \\
&&& chaos for  & or chaos & or chaos \\
&&& f/f$_0$ $\sim$ 0.66 - 2.2 & &\\

NTE610 & 45 & 16 & Period doubling and  & Period doubling & No period doubling \\
&&& chaos for  & $\underline{only}$ for  & or chaos \\
&&& f/f$_0$ $\sim$ 0.14 - 3.84 & f/f$_0$ $\sim$ 1.17 - 3.25 &\\

\end{tabular}  

\end{table}

\end{document}